*Review*

# Objective Diagnosis for Histopathological Images Based on Machine Learning Techniques: Classical Approaches and New Trends

**Naira Elazab** [1]**, Hassan Soliman** [1]**, Shaker El-Sappagh** [2,3]**, S. M. Riazul Islam** [4,*] **and Mohammed Elmogy** [1]

[1] Information Technology Department, Faculty of Computers and Information, Mansoura University, Mansoura 35516, Egypt; naira.elazab@mans.edu.eg (N.E.); hsoliman@mans.edu.eg (H.S.); melmogy@mans.edu.eg (M.E.)

[2] Centro Singular de Investigación en Tecnoloxías Intelixentes (CiTIUS), Universidade de Santiago de Compostela, Santiago de Compostela 15705, Spain; shaker.elsappagh@usc.es

[3] Information Systems Department, Faculty of Computers and Artificial Intelligence, Benha University, Benha 13512, Egypt

[4] Department of Computer Science and Engineering, Sejong University, Seoul 05006, Korea

* Correspondence: riaz@sejong.ac.kr; Tel.: +82-02-3408-2969

**Abstract:** Histopathology refers to the examination by a pathologist of biopsy samples. Histopathology images are captured by a microscope to locate, examine, and classify many diseases, such as different cancer types. They provide a detailed view of different types of diseases and their tissue status. These images are an essential resource with which to define biological compositions or analyze cell and tissue structures. This imaging modality is very important for diagnostic applications. The analysis of histopathology images is a prolific and relevant research area supporting disease diagnosis. In this paper, the challenges of histopathology image analysis are evaluated. An extensive review of conventional and deep learning techniques which have been applied in histological image analyses is presented. This review summarizes many current datasets and highlights important challenges and constraints with recent deep learning techniques, alongside possible future research avenues. Despite the progress made in this research area so far, it is still a significant area of open research because of the variety of imaging techniques and disease-specific characteristics.

## 1. Introduction

Medical Images are a fundamental section of each patient's digital health file. Such images are produced by individual radiologists who are restricted by speed, professional weaknesses, or a lack of practice. It requires decades and reasonable financial resources to train a radiologist. Additionally, some medical care methods outsource radiology confirmations to less economically developed nations, such as India, via teleradiology. A late or incorrect analysis can cause injury to the patient. Thus, it would be beneficial for medical imaging (MI) analyses to be performed by automatic, precise, and effective machine learning (ML) algorithms. MI analysis is a significant research area for ML, in part because the information is somewhat organized and labeled; i.e., this is probable if the patient was examined in a region with good ML systems [1]. That is significant for two reasons. First, with regards to real patient metrics, MI analysis is a litmus check regarding whether ML techniques would, in actuality, improve individual outcomes and survival. Second, it provides a testbed for human–ML

interactions—i.e., how responsive is an individual likely to be to the health changing possibilities being put forward or aided by a nonhuman actor [2]. In recent years, ML has shown significant advances. For a wide variety of applications, including image recognition, medical diagnosis, defect identification and construction health assessments, the potential of this field has also expanded. These new developments in ML are due to many factors, like the creation of self-learning mathematical models that enable computer techniques to execute particular (human-like) tasks based solely on learned patterns, in addition to the increase in the computer power that supports these models' analytical capabilities [3].

There are many imaging types, and their use is becoming more widespread. Types of MI include ultrasound, X-ray, magnetic resonance imaging (MRI), retinal scans, histopathology images (HI), computed tomography (CT), positron emission tomography (PET), and dermoscopy images. Some examples of MIs are shown in Figure 1. Many of these types analyze numerous organs, such as CT and MRI, whereas others are organ-specific, such as retinal and dermoscopy images [4]. The quantity of produced information from each analysis stage differs depending on nature of the MI and the tested organs. HIs are useful for biological studies and to make medical decisions. In addition, they are generally utilized to provide "ground truths" (GTs) for other modalities of MI, such as MRI. A histology slide is a digital record a few megabytes in size, while a magnetic resonance image can be several hundred megabytes. This has a technical effect on how the data is preprocessed and on the architecture design of the algorithm in terms of processor and storage limitations [5].

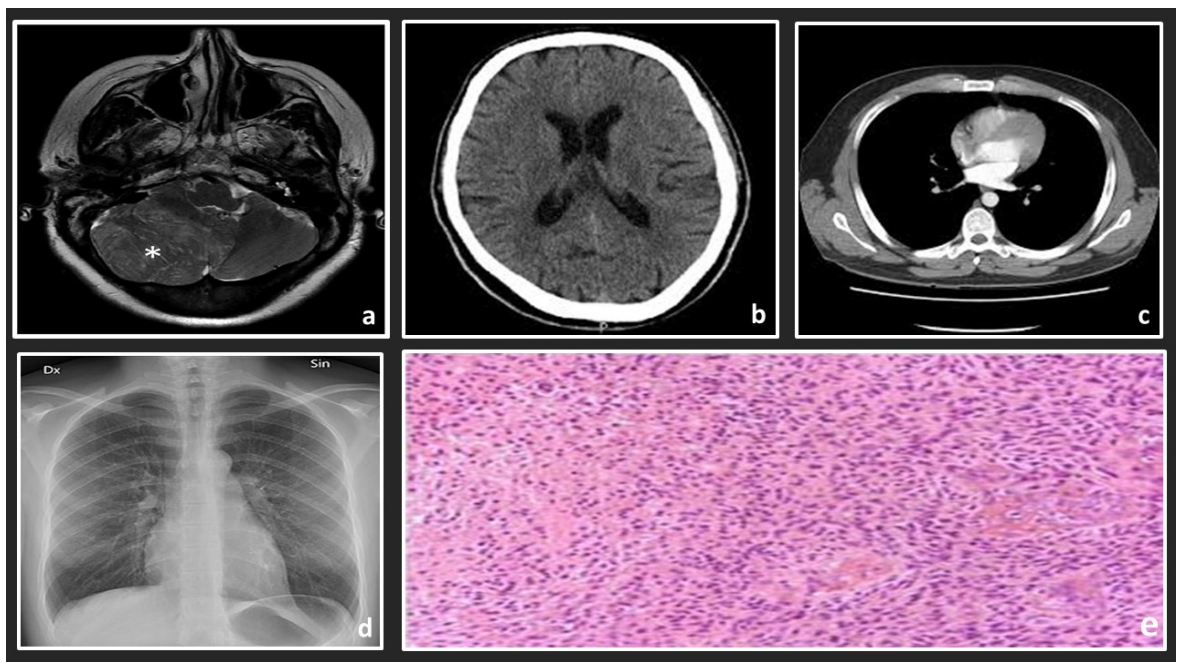

**Figure 1.** Examples of some medical image types: (**a**) MRI scan of the left side of a brain; (**b**) an axial CT brain scan; (**c**) an axial CT lung scan; (**d**) chest x-ray; (**e**) a histology slide with high-grade glioma.

Pathology analyses are traditionally executed by an individual pathologist observing a dyed specimen on a glass slide with a microscope. Lately, efforts have been made to record the whole slide with a reader and save it as an electronic picture, called a whole slide image (WSI) [6].

Digitizing pathology is just one recent development that produces high levels of visible information designed for automated diagnoses. It enables us to see and understand pathologic cell and muscle samples in good quality images with assistance from personal computer tools. It also brings about the possibility of applying image analysis techniques. Such techniques would assist pathologists and support their explanations, such as hosting and grading. Various classification and segmentation methods for HI have already been discussed in this review. We present and compare conventional techniques and deep learning (DL) methods to choose the most appropriate method for histopathology issues [7].

Natural microscopic architecture data and their features at nuclei, tissue, and different organ levels could be key to illness expansion and infection treatment analysis. Additionally, to examine and diagnose the histological image of biologic microscopic, pathologists have identified the morphological features of tissue that show the current presence of infection, such as cancer [8].

Some characteristics of disease, such as tumor-infiltrating lymphocytes, might be deduced from HI alone. Additionally, HI analysis, which is called the "gold standard" in many disease diagnoses, is nearly included in all kinds of cancer detection and treatment procedures. HI needs specific analysis with respect to organs and a specific task for the visualization of various tissue components under a microscope. With one or more stains, the sections are dyed. These are staining attempts to uncover cellular elements. The contrast is shown by using counterstains [9].

Efficient ML algorithms are presented and used in HI analysis to help pathologists to acquire a quick, stable, and quantified examination result for a more accurate diagnosis. Many different traditional and deep learning methods support the pathologists in accessing more tissues to determine the internal relationship between the visual images and the specific illness. Additionally, since the ML techniques are generally semi- or fully automated, they are effective, encouraging technical feasibility for histopathology examination within the recent big data age [10].

On the other hand, most of the HI analysis stages are based on mathematical basics. Mathematical operations and functions are applied to all analysis stages, starting from the preprocessing to diagnosis stages to provide an intensive analysis for HIs. Figure 2 illustrates the main phases of a common histopathological images pipeline based on conventional ML techniques. First, HIs are supplied to the system as a 2D array for grayscale images or a 3D array for colored images. Then, the preprocessing stage applies some linear algebra operations on the image array to enhance the image quality. This stage helps to distinguish significant structures from others in the processed images. Third, the segmentation stage is applied to differentiate the cells from other background objects by applying some state-of-the-art mathematical algorithms, such as thresholding, level set, watershed transform, and intensity and texture homogeneity transforms. Fourth, the feature extraction stage extracts the most significant features in the segmented images instead of processing each pixel, which reduces the system's computation complexity. Besides, most handcrafted features are based on applying some mathematical techniques to detect the changes in the intensity, color, or texture of the pixels. Common derivative techniques are utilized to detect these changes by applying first or second derivatives to pixel values. Finally, the diagnosis stage is applied to classify or cluster the processed images, depending on the extracted features. The classification and clustering techniques are based on applying some mathematical operations that distinguish the processed images based on the extracted features.

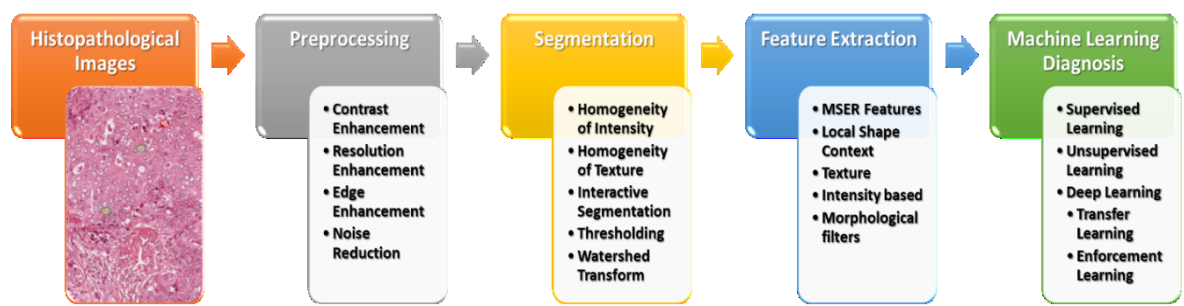

**Figure 2.** An overview of the HI analysis pipeline.

Numerous segmentation and classification techniques for primitives tissue in HIs were presented in this respect. For selecting the appropriate HIs analysis method, the various ML methods on HIs are reviewed in this study. In this work, the digital HI analysis, applying different ML algorithms and their issues, are described. The paper is presented to describe the necessity of the analysis procedure for segmentation and classification in computer-aided diagnosis (CAD) systems using HI.

The rest of the article is organized into six sections. Section 2 gives a brief overview of the fundamental histopathological analysis. In Section 3, the conventional approaches for HI analysis are described. Section 4 introduces the use of deep learning techniques in HI analysis. The datasets, the discussion, and tasks of HI analysis are elucidated in Sections 5 and 6, respectively. Limitations and future trends in the HI analysis are introduced in Section 7.

**2. Histopathological Image Overview**

HI has natural and abnormal biological structures, as well as morphological and architectural features defined by pathologists, based on their knowledge. Even given the tissue area, some structures are small, and related patterns typically have high visual appearance variability. In biological systems and anatomy, most visual variability is inherent [11].

Next to obtaining electronic HI via the biopsy test, the guide analysis of images contributes to variability in diagnosis and treatment. To get over this issue, CAD techniques are applied to provide an objective examination of disease. The fundamental steps necessary for applying the CAD examination system appear in Figure 2. This includes electronic image handling methods, such as segmentation, feature extraction, and classification [12].

HI analysis contains the computations executed at various zoom scales (×2, ×4.5, ×10, ×20, and ×40) for multivariate mathematical examination, analysis, and classification. It could be achieved at a lower zoom for tissue stage examination. Demir et al. [13] presented tissue stage and cell stage examination techniques for cancer diagnosis. They examined HI by applying preprocessing, feature extraction, and classification strategies. The new improvement in electronic pathology requirements for the growth of quantitative and automatic digital image examination methods aids pathologists in understanding the number of digitized HIs [14].

*2.1. Types of ML Systems in HI Analysis*

2.1.1. Computer-Aided Diagnosis (CAD)

Many of the searched tasks in electronic HI analysis are CAD systems, which are the pathologists' fundamental functions. The diagnostic method includes the function to map WSI to one of many infection types, indicating a supervised learning function. Considering that the mistakes created by the ML process vary from those created by an individual pathologist [15], the enhancement of classification reliability would be increased by applying the CAD method. CAD could also reduce the instability in understanding and reduce overlooking by analyzing each pixel in WSI. Different related diagnosis functions contain the recognition or segmentation of the region of interest (ROI), such as tumor area in WSI [16,17], rating of immunostaining [18,19], cancer phase [20,21], mitosis recognition [22,23], gland segmentation [24–26], and quantification of general intrusion [27].

2.1.2. Content-based on Image Retrieval (CBIR)

CBIR retrieves pictures related to a query picture. In electronic pathology, CBIR methods help in several scenarios, especially in examining, training, and studying [28–30]. For example, CBIR methods could be utilized for academic applications and novice pathologists to recover appropriate instances of HI of tissues. Additionally, such methods would also be useful to skilled pathologists, especially while detecting uncommon cases. Because CBIR certainly does not need tag data, unsupervised learning could be utilized [31]. Not just precision, but additionally high-speed research of related pictures from several pictures are needed in CBIR. Thus, numerous approaches can reduce picture feature dimensionality—such as primary element examination and small bilinear combination [32]—and quickly estimated the closest neighbor searches [33].

2.1.3. Finding New Clinicopathological Associations

Traditionally, several essential discoveries regarding diseases, such as tumor and contagious conditions, have already been produced by pathologists and analysts. They cautiously and carefully examined pathological specimens. For example, pathologists analyzed the gastric mucosa of individuals with gastritis in [34]. Efforts were made to link the morphological options that come with cancers using their medical behavior. For instance, tumor grading is essential in a patient's diagnosis and in preparing treatment for many kinds of cancer, such as breast and prostate cancer.

There is a noticeable development in the digitization of clinical data, which later improved the genome evaluation technique. Therefore, a wide range of electronic data, such as genome data, electronic pathological images, MRI, and CT scans, are now accessible [35]. By examining the connection between these imaging modalities, new hospital pathological associations, such as the connection involving morphological quality and somatic cancer mutation, are available [36]. CAD techniques could be subdivided into conventional ML and DL methods, illustrated in more detail in the next few sections.

## 3. Conventional Machine Learning Methods

CAD systems played an essential role and have become an important research topic in HI and diagnostics. Various image processing techniques were applied to examine the disease's diagnosis and prognosis for these HIs. Various image processing and computer vision (CV) techniques have been implemented for gland and nuclei segmentation, cell kind recognition, or classification to extract quantitative measurements of disease characteristics from HIs and automatically assess whether or not a disease exists inside examined samples. It could help to determine the degree of seriousness of the disease, whether present in the sample. Conventional ML methods often contain a few steps to manage HI, as shown in Figure 3. Each step is illustrated in the following sections.

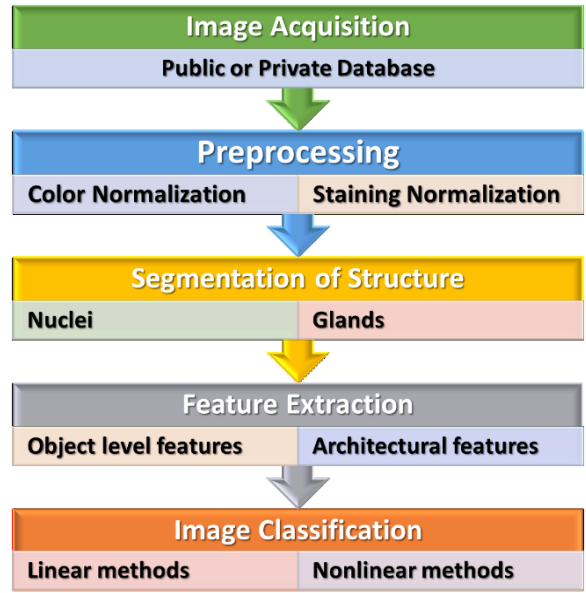

**Figure 3.** The conventional machine learning methods for HI.

*3.1. Preprocessing*

Preprocessing could recompense for variations between images, which can vary in color, staining, and other problems, such as noise, which are usually due to the scanning procedure. The gross sections are made with wax to analyze the tissue's architecture and components under the microscope and colored with one or more stains. Pathologists use staining to isolate cellular components for the diagnosis of structural as well as architectural tissue analysis. Hematoxylin–Eosin (H&E) staining is most commonly utilized, and it separates the connective tissue, cytoplasm, and nuclei. Nuclei are stained blue by Hematoxylin, while connective tissue and cytoplasm are stained pink by Eosin. DAB, immune-histochemistry stains, etc. are the other stains. The consistency of the features extracted from the image directly affects classification performance. Thus, it is essential to define the proper conditions under which the image preprocessing techniques will work as the first job. Noise and various illumination fluctuations are detrimental to image processing techniques. If those negative factors are eliminated, it will improve performance. Pre-processing imaging techniques are well adapted to this mission. Pre-processing methods control changes in image brightness and contrast and eliminate noise.

3.1.1. Staining Normalization

HI could have powerful color variations due to various scanners, various staining techniques, and sample age. An efficient color calibration between samples is difficult to accomplish [37]. Hence, color normalization is needed in most of the processing scenarios. Deconvolution-based methods and histogram-based methods are examples of color normalization [38]. Anghel et al. [39] suggested improving stain normalization in low-quality WSIs to increase ML pipeline accuracy. They used an ML pipeline based on convolutional neural networks (CNNs), which classifies pictures to detect prostate cancer, to demonstrate the robustness of this new normalization process. This system makes it possible to pre-process massive datasets and is a crucial requirement for any biomedical imagery learning computer.

3.1.2. Color Normalization

Color normalization is required for bright and fluorescent HI analysis. This process decreases the variations in samples of tissue because of variance in conditions of scanning and staining. There are different techniques for the color normalization of HI, such as the Reinhard approach, descriptor of stain color, and histogram specification [40]. For HI research, MIAQuant [41] was stained by different approaches and obtained with various instruments. The machine automatically extracts and quantifies markers with various colors and types and, for the visual comparison of their positions, aligns the contiguous tissue slices, stained by multiple markers. MIAQuant segments markers efficiently and quantifies them by integrating clear and effective imaging techniques with precise colors from histological images. MIAQuant aligns and measures a picture with contiguous (serialized) parts of the cloth, where the markers are covered with different colors so that the markers can be visually comparable. Its successful findings in biomedicine have inspired us to increase the capacity to communicate different marker positions and, finally, neighboring serialized pieces of tissue. Easy, efficient, and effective processing, pattern detection, and supervised teaching techniques with their improved framework, called MIAQuant-Learn [42], enable you to personalize marker segmentation by all colors.

The quality of HI is the parameter to determine that it can be the most remarkable approach for color normalization. Metrics of quality, such as the structural similarity index metric (SSIM), contains three factors (contrast, luminance, and structural) [43]. These parameters are given in Equations (1)–(3), respectively.

$$N(x, y) = \frac{2\ \sigma_x\ \sigma_y + c_2}{\sigma_x^2 + \sigma_y^2 + c_2} \quad (1)$$

$$M(x, y) = \frac{2\ \overline{X}\ \overline{Y} + c_1}{\overline{X}^2 + \overline{Y}^2 + c_1^2} \quad (2)$$

$$R(x, y) = \frac{\sigma_{xy} + c_3}{\sigma_x\ \sigma_y + c_3} \quad (3)$$

where $\overline{X}$ and $\overline{Y}$ are means of origin and processed image, respectively. $\sigma_x$ and $\sigma_y$ are standard deviation, $\sigma_{xy}$ is the correlation coefficient between the processed and the source image. $c_1$, $c_2$, and $c_3$ are constants that could stabilize SSIM if nearing a zero value. By using Equations (1)–(3), the SSIM is derived from Equation (4). The SSIM value is 0 to 1. The better approach is color normalization, when the value is near to 1.

$$\text{SSIM} = \left(\frac{2\ \overline{X}\ \overline{Y} + c_1}{\overline{X}^2 + \overline{Y}^2 + c_1^2}\right)\left(\frac{2\ \sigma_{xy} + c_2}{\sigma_x^2 + \sigma_y^2 + c_2}\right) \quad (4)$$

*3.2. Recognition and Segmentation of Structures*

One of the main tasks in HI analysis is image segmentation, and it has been applied to solve a wide variety of issues. Image segmentation, in its entirety, similar to clustering, is an unplaced issue as defining a meaningful segment can vary from task to task or even from image to image. For this purpose, the application domain must be aware of the segmentation algorithms, either by taking custom features or algorithmic methods into account or by learning from vast volumes of knowledge [44].

The existence of pathology and the number and the morphological features of detailed textures, such as nuclei and glans, are essential variables to analyze the existence and intensity of pathology—for example, colorectal [45], prostate [46], and breast [47] cancer.

3.2.1. Nuclei and Cells

Nuclei would be the main organelles of a eukaryotic cell, comprising the majority of cell DNA. Nuclei examination often requires recognition, segmentation, and separating overlaps. Recognition of seed factors in nuclei is needed by several segmentation and checking techniques [48]. Several methods had been proposed in the review for nuclei recognition, involving techniques predicted on Euclidean range chart peaks [49], Hough change (recognizing seed factors for circular formed textures, requesting extensive computation) [50], Laplacian of Gaussian filters [51], and radial symmetry [52]. Several methods were shown to represent accurate segmentation. The techniques based on thresholding and morphological procedures are appropriate on a standard background [53]. They may not, however, be powerful in measurement, form, and structure change. Effective shape forms could mix picture attributes with nuclei form types [54]. However, they depend on seed factors. Different techniques were predicated on gradients in polar [55] and graph reductions [56]. Ta et al. [53] suggested an approach, dependent on regularization graphs. The method's specificity was to utilize graphs as image confidential modeling at various grades (areas or pixels) and various component relations, such as grid graph. Dependent on Voronoi's diagrams, they suggested a graph reduction technique for nucleus segmentation of HIs for serious cytologic and breast cancer. A pseudo metric $\delta: V \times V \to R$ is illustrated as

$$\delta(u,v) = \min_{\rho \in P_G(u,v)} \sum_{i=1}^{m-1} \sqrt{w(u_i, u_i+1)}\ (f(u_i+1) - f(u_i)) \quad (5)$$

where a weight function between two pixels is w ($u_i$, $u_i$ + 1), and a set of paths connecting two vertices is PG (u, v). Taking into account the set of K seeds $S = (s_i \subseteq V)$, where i = 1, 2, ..., K, the energy $\delta: V \to R$ presented the metric $\delta$ for all the seeds of S, which can be presented as:

$$\delta_s(u) = \min_{s\ i \in S} \delta(s_i, u), \qquad \forall_u \in V \quad (6)$$

The zone z of control (known as the Voronoi cell) of the seed $s_i \in S$ is the set of vertices nearer to $s_i$ than any other seeds related to the metric $\delta$. It can be defined, $\forall j = 1, 2, ..., K$ and $j \neq i$, as

$$z(s_i) = u\ \in V:\ \delta(s_i, u) \leq\ \delta(s_j, u) \quad (7)$$

Then, the energy distribution of the graph is the set of powerful zones Z (S, $\delta$) = {Z ($s_i$), $\forall\ s_i \in\ S$}, for a given set of seeds S and a metric $\delta$.

3.2.2. Glands

It is organs that are shaped by an ingrowth from the epithelial surface. Techniques in thresholding and area growing could recognize nuclei and lumen, which can be applied to initial seed factors for area growing [57]. Segmentation predicated on polar ordinates (the middle of the gland) was performed on the benign and malign gland [48].

Rittscher et al. [58] caught some bright pixels participating in a standard distribution form. Their technique applied three features. The first was the intensity of fluorescent emission. The others were derived from curve descriptors, which could be calculated from eigenvalues of the Hessian matrix. The eigenvalues ($\lambda 1(x, y) \le \lambda 2(x, y)$) of the image $I(x, y)$ encode the curve data of the image. They give helpful cues for shape detection, such as structures of the membrane. However, the eigenvalues are influenced by the brightness of the image [6]. Equations (8) and (9) represent two features, which are autonomous from the image's brightness. These features are known as normalized-curve index and shape index, respectively.

$$\phi(x, y) = \tan^{-1} \frac{\sqrt{(\lambda_1 (x,y))^2 + \lambda_2 (x,y)^2}}{I(x,y)} \tag{8}$$

$$\theta(x, y) = \tan 2(\lambda_1 (x,y), \lambda_2 (x,y)) \tag{9}$$

This segmentation, based on normalized-curve index and shape index divides an image's pixels into three sets: foreground, indefinite, and background. The indefinite set covers all of the pixels which are not involved in the other two sets. From these sets, the intensity distribution of foreground and background and log-likelihood intensity are derived.

*3.3. Feature Extraction*

HI was examined by applying several descriptors based on the information of domain experts. Analysis requirements are represented primarily with cytological phrases (i.e., glands, nuclei) and its involvement in the malignant and benign surface. Consequently, many papers dealt with the object stage (applying segmented item attributes) and object connection stage (applying structural attributes). Tumors, such as ductal carcinoma and lobular carcinoma, have an abnormal growth of epithelial cells in these structures. The abnormal growth of tissue, representing a tumor, may result in a large number of nucleus cells or a high number of mitotic cells in a small area. HI captures this function, but it captures other healthy tissues in addition to the nucleus, which can be seen in images of benign tumors. Stroma is a kind of tissue that can be seen in malignant and benign images with the same characteristics. The classification method could be enhanced by choosing more appropriate patches. Histopathological considerations remain paramount in this regard. There are well-known considerations, such as the size of the tumor, the histological shape and subtype, the nature of the sign, circular morphology and degree of differentiation, and the presence of vascular lymph invasion and the involvement of lymph nodes. We have gone from a greater understanding of these causes in recent years, identifying significant factors, such as tumor budding and lymphocytic infiltration. The prognostic importance of resection margins has also been assessed over the last two decades—particularly circumferential margins. Some patients are also notable with histological features associated with various molecular and genetic markers, including KRAS, BRAF, and microsatellite instability. The feature can be divided into literature level artifacts and structural features. Object-level characteristics are characteristics correlated with the nucleus size and shape. Features of the structure describe topological characteristics based on graph theory.

*3.3.1. Object-Level Characteristics*

Object-level characteristics rely firmly on regarded items (often gland or nuclei) and segmentation methods [59]. These characteristics are appropriate at any resolution. However, generally, they are produced by high-resolution pictures. Object-level characteristics were generally produced to each shade channel and could be collected into shape characteristics, such as region [60]. In the pre-segmentation procedure based on the unregulated medium-shift cluster, Kuse et al. [61] applied a feature extraction method. Thresholds limit the color variation to the image section. After this stage, the kernel is formed, and the contour and area constraints are removed from the overlap. Gray-level co-occurrence Matrix (GLCM) texture functions are finally extracted from a sectioned picture utilizing a classifier specified by support vector machines (SVM). Caicedo et al. [24] merged seven approaches of extraction of features and construct a kernel-based representation of data for each form of function. Inside the SVM classifier, kernels are used to find similarities between data and enforce a content retrieval mechanism.

3.3.2. Structural Characteristics

Structural characteristics are primarily based on graphs. They are comprised of nodes that can be linked by arcs. Lately, characteristics based on the graph were investigated frequently because they are suitable to characterize tumor structure. Three types of brain cancer classification (inflame, health, maligning) were conducted by Demir et al. [13]. They utilized a total weighted graph. Chekkoury et al. [62] made a hybrid between features based on texton and morphologic system to breast cancer. Doyle et al. [63] applied a mix of characteristics for prostate cancer. They show 90% precision in characterizing cancer and benign tissue. To handle the multi-resolution properties of HIs and emulate pathologists' approach to analytics, multi-resolution methods have been proposed. In many resolutions, the Gaussian pyramid method portrayed the pictures. Features for every level have been extracted separately and labeled as image tiles. Color and texture characteristics are typically utilized at low resolutions. The medium-scale architectural arrangement of glands and nuclei at high resolutions may be discriminatory.

3.4. Classification

Techniques for classifying their goal determine the group of recent observations between some classes based on marked training group. Regarding the function, anatomical composition, characteristics, and the preparation of tissue, the classification differs. DiFranco et al. [64] classified the prostate tissue into seven classes: benign hyperplasia, inflammation, Gleason rank 3 and 4, and intraepithelial neoplasia. They acquired 83% accuracy based on sum average, contrast, connected histogram characteristics, and entropy.

Huang et al. [65] proposed a method to enhance hepatocellular carcinoma classifying, which used a subset of feature selection with a support vector machine based on a decision graph for every decision node. Alexandratou et al. [66] presented a literature review for prostate cancer, illustrating good cancer recognition. The overview of conventional HI analysis methods is summarized in Table 1.

**Table 1.** Overview of HI analysis for different conventional methods.

| Study | Organ | Method | Results | Problem with Method |
|---|---|---|---|---|
| Basavanhallya et al. [67] | Breast | Hierarchical Normalized cut | 89% accuracy of segmentation | Discovers false-positive errors because of lumen presence. |
| Khadi [68] | Meningioma | Classification of texture applying fractal features | 92.5% accuracy applying individual texture measurement to meningioma tissue | Misclassification results because of the non-uniformity cell construct. |
| Demir et al. [69] | Colon | Object graph approach for segmentation | 87.59% accuracy of compatible images | Requires variable optimization that reduced results for segmentation |
| Tosum et al. [70] | Breast | Diagram run-length models to segmentation of the image | The novel descriptor of texture for unsupervised classification had 99.0% accuracy for gland segmentation | Complexity relies on the number of primitives in picture |
| Chekkoury et al. [62] | Breast | A novel hybrid between features based on texton and morphologic system | 86% classification accuracy combining textural features and morphometric | The effects of image compression on classification accuracy |
| Doyle et al. [63] | Prostate | mix of characteristics | 90% accuracy in characterizing cancer and benign tissue | Limited Feature Set |
| DiFranco et al. [64] | Prostate | classified into seven classes (benign hyperplasia, inflammation, Gleason rank 3,4, and intraepithelial neoplasia) | 83% median accuracy based on some characteristics like sum average, contrast | The computation time required for reading and writing data to and from disk, particularly in feature extraction |

## 4. Deep Learning Methods

Recently, DL techniques have often been studied in the effective form of ML methods. Within the last few years, DL techniques outperformed traditional ML methods in varied fields, such as CV, natural language processing (NLP), biomedical fields, and automated analysis for HI [71]. DL methods in the CV are derived from the structure levels for nonlinear transformations on natural input pixels. This structure formed significantly abstract representations, which could be realized in a hierarchical style [72]. A typical instance of a commonly applied structure is the CNN [73].

Multiple criteria can be considered when using the DL techniques to deal with histopathology, since accomplishing the method is partly due to the task-species setting. Among the principal features of HI is that appropriate styles be determined by the magnification stage. The key factors are the size of the patch given to the network, the localization of parts in the image where appropriate histopathology originals can be found, and the homogeneity of staining for WSI [74]. The network structure represents an important position, while many studies keep predefined system structures, as illustrated in Figure 4.

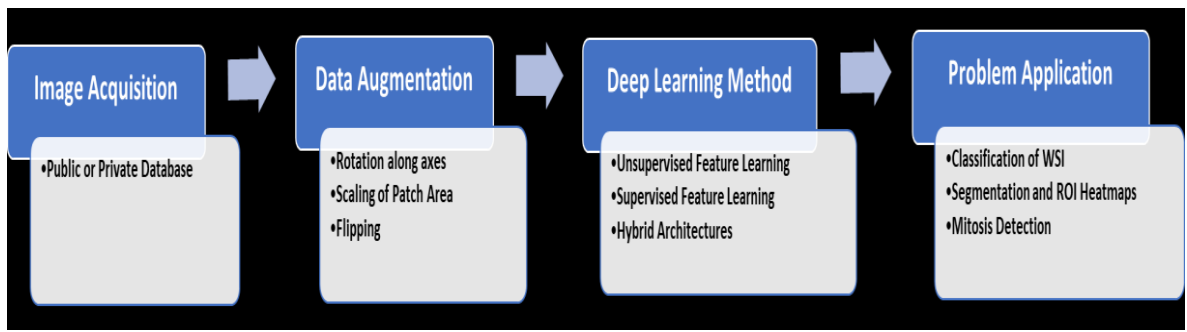

**Figure 4.** The typical deep learning steps for HI analysis.

The majority of the DL techniques for localizing, classifying, and segmenting HI are somewhat recent. Deep neural techniques are stated in the new literature of HI analysis, such as [59,13,75]. For example, Irshad et al. [48] were the first mentioned in a review. The critical patterns from an exhaustive analysis of different nuclei identification, segmentation, and classification approaches utilized in HI, specifically in H&E staining protocols, were described and discussed in this review. Ciresan et al. [56] presented one of the first significant efforts to utilize the deep method in mitosis recognition for HI analysis. Arevalo et al. [76] presented a hybrid illustration method to the basal cell of carcinoma areas and utilized a topographic unsupervised technique and a case of characteristic illustrations. They increased the classifier's efficiency by 6% regarding traditional structure-based discrete cosine transform (DCT). Nayak et al. [77] presented an alternative method for the unsupervised Boltzmann technique for understanding image signatures. They classified images of the cancer genome atlas (TCGA) for apparent cell-kidney cancer and glioblastoma variform. The last stage was created utilizing the classifier of multi-class support vector machines (SVM) techniques.

Malon et al. [78] proposed a novel mix of features. The authors compiled a feature set of the fundamental nucleus and cytoplasm pixel statistical measurements and combined them with a CNN classifier. Their method increased the efficiency when comparing to the characteristics of handcrafted methods. Xu et al. [79] created an approach that handled marked cases. Numerous examples of the learning platform were presented in this technique, where the colon HI classifier was developed. The researchers presented a thoroughly supervised method and also weakly-marked one. Hou et al. [80] presented a similar strategy by applying numerous cases to understand how to categorize low-grade glioma and glioblastoma images in TCGA. The technique used three phases. First, it understood the masks of discriminative parts utilizing CNN form with few picked discriminative areas. It then created a patch level forecast, applying CNN. Finally, the counting of the class was generated.

Arevalo et al. [81] proposed a stacked form that revealed the most significant features when mixing characteristics from two-layered topographic independent component analysis (TICA) around patches for finding basal cell cancer. They presented an electronic discoloration technique because characteristic detectors are weighed with classification likelihood to spotlight parts, which can be most linked to carcinoma. This technique accomplished 99% of the area under the curve (AUC) for 100,000 patches. Hang et al. [82] were dependent on the understanding book of 1024 characteristics to classify apparent cell cancer and also glioblastoma multiforme. The book was constructed with a stacked unsupervised technique utilized in the spatial chart corresponding platform with the SVM classifier's final step. Some new studies were interested in the classifying of the H&E patch. Han et al. [83] provided a novel deep unsupervised technique for glioblastoma multiforme characterization. They can distinguish two critical phenotypic subtypes at various survival shapes, utilizing the produced unsupervised characteristics. Noel et al. [84] applied a group of 30,000 patches produced from the WSI of the International Conference on Pattern Recognition (ICPR) contest to recognize breast carcinoma, categorizing every pixel utilizing CNN mitosis, stoma, and lymphocytes. They reached 90% accuracy, indicating a better WSI primitive classifier, which could enhance classification performance.

Romo-Bucheli et al. [85] served a CNN form with prospect patches for calculating tubule associateship likelihood to measure tubule nuclei, which was related to high–low chance classes decided by the Oncotype DX test. DL stains' application distinctive from H&E continues to be not adequately researched, such as immunohistochemistry. Chen et al. [86] presented the recognition of immune cells using seven-layer CNN with areas of nature colors from RGB channels, displaying markers of immune cells. They compared the efficiency of their recognition method with pathologists' efficiency, attaining a 99% correlation coefficient. A review of deep-neural models, developed for HI analysis, was presented by Srinidhi et al. [87].

Sumi et al. [88] proposed a deep spatial fusion network that manages the dynamic construction of discriminative characteristics over patches. In the high-resolution histology picture, it also learns to change the bias on patch-wise prediction. To extract characteristics from the cellular level to the tissue level, Patch wise InceptionResNetV2 is used. This approach is used to analyze the spatial interaction between patches. Compared to previous CNN experiments using different architectures, better performance is given by their proposed system. This work needs to be expanded to include other networks that effectively examine more malignant tumor types other than glioblastoma and oligodendroglioma. Maximal tumor resection is especially necessary for larynx surgery, thus maintaining adjacent healthy tissue. Therefore, accurate and swift intraoperative laryngeal histology is vital for optimal surgery results. Zhang et al. [89] hypothesized a DL stimulation of Raman scattering microscopy (SRS) that could automatically and precisely diagnose new, unprocessed surgical specimens with laryngeal squamous cell cancer without fastening, separation or staining. First, they compared 80 pairs of adjacent SRS and regular frozen parts to determine their concordance. They then applied SRS imaging to 45 patients' fresh chirurgical tissues based upon a DL model for automatically producing histologic results. They also applied the main diagnostic features.

Pathology's scientific function is to diagnose diseases to classify differences at cell structures level, such as nucleolus and cytoplasm, tissue (i.e., cell community with complicated structures), and organs that give rise to patient symptoms. It was found that damaged or unresolved cells do not die, and uncontrolled growth is seen by clinical pathology framework and histopathology methods, explaining cancer cells' mass production. Cancer cells also migrate through the blood and lymph systems and cross borders to another body area, reproducing the uncontrolled growth cycle method. This cancer cell phase is called metastatic spreading or metastasis that leaves one area and grows in another part of the body. Breast cancer could be detected by a histopathological methodology. This diagnosis can be made using different ML Models, and DL-based CNN Models. Agarwal et al. [90] showed that CNN models provided significant accurate results in the comparison of ML models, such as U-Net [91], improved precision, and high-performance segmentation. In conjunction with very low-cost consumer graphics processing units, large images can therefore be processed rapidly.

Approaches that rely on Generative Adversarial Networks (GANs) are likely to minimize the need for large volumes of manual notations. Not only have recent innovations enhanced initiatives but so have new technologies. Now, unattended techniques may carry out various tasks for which supervised methods are indispensable. The latest state-of-the-art advances in histopathological images of GANs were summarized in [92]. The overview of the discussed studies is summarized in Table 2.

**Table 2.** Overview of supervised and unsupervised learning models based on DL techniques.

| Study | Organ | Staining | Potential Usage | Method |
| --- | --- | --- | --- | --- |
| **Supervised Learning** | | | | |
| Litjens et al. [93] | different tissue | H&E | Prostate and breast carcinoma detection | Convolutional Neural Network based on pixel classifier |
| Nagpal et al. [94] | Prostate | H&E | Anticipating Gleason indicator | CNN based on sectional Gleason model classifier + k-nearst neighbors (KNN) based on Gleason grade anticipation |
| Zhao et al. [95] | Breast | H&E | Metastasis Detection + classification | Characteristic pyramid collecting based on the fully convolutional network (FCN) system with the synergistic training technique |
| Xing et al. [96] | different tissue | H&E, Immunohistochemistry (IHC) | Segmentation of nuclei | CNN + selection based on sparse form Pattern |

| Reference | Organ | Stain | Task | Method |
|---|---|---|---|---|
| Gu et al. [97] | Breast | H&E | Tumor detection | U-Net based on multiple resolution model with multiple encoders and a singular decoder system |
| Tellez et al. [98] | Breast | H&E | Detection of Mitosis | Train of Convolutional Network applying H&E registered to PHH3 slides as a reference |
| Wei et al. [99] | Lung | H&E | Histological subtypes of lung gland classifier | ResNet-18 on the basis of patch classification |
| Song et al. [100] | Cervix | Papanicolaou (Pap), H&E | Cells Segmentation | Multiple level CNN system |
| Agarwalla et al. [101] | Breast | H&E | Segmentation of tumor | CNN and 2D- Long short-term memory (LSTM) to representing training and context collecting |
| Ding et al. [102] | Colon | H&E | Glands segmentation | Multiple level FCN network with a high-resolution section to avoid the lost in highest pooling layers |
| Bejnordi et al. [103] | Breast | H&E | Invasive Carcinoma detection | Multiple level CNN which first determines tumor-associated stromal modifications and more categorize into normal/benign versus invasive carcinoma |
| Seth et al. [104] | Breast | H&E | Ductal carcinoma in-situ (DCIS) segmentation | Compared UNets learned in many resolutions |
| **Unsupervised Learning** | | | | |
| Xu et al. [105] | Breast | H&E | Segmentation of nuclei | Stacked sparse autoencoders |
| Bulten and Litjens [106] | Prostate | H&E | Tumor classification | Convolutional adversarial Autoencoders |
| Hou et al. [107] | Breast | H&E | Segmentation and detection of nuclei | Sparse autoencoder |
| Sari and Gunduz-Demir [108] | Colon | H&E | Feature extraction and classification | Restricted Boltzmann + clustering |
| Gadermayr et al. [109] | Kidney | Stain agnostic | Object of interest segmentation in WSIs | CycleGAN + UNet segmentation |

| Gadermayr et al. [110] | Kidney | Periodic acid–Schiff (PAS), H&E | Glomeruli segmentation | CycleGAN |

## 5. Datasets

The size of the datasets given to researchers for training and testing their methods has dramatically increased in the latest challenges. There is a set of public databases in the electronic pathology subject that include manual annotations for HI, as listed in Tables 3 and 4 [111]. They might help the examination objectively. Slide issue (stain) and image issue (image resolution, zoom level) are similar. However, all these databases are targeted to specific diseases. These databases do not handle several tasks. Additionally, there are many high scale HI datasets, which include WSIs of high resolutions.

TCGA [33] includes around 10,000 images from different types of cancer. Genotype-Tissue Expression (GTE) [116] includes around 20,000 WSIs from different tissues. The Stanford Tissue Microarray Database (TMAD) is available for researchers to access images of microarrays for tissue. It provides images of archiving 349 distinguished probes on 1488 microarray slides of tissue [113]. The CAMELYON dataset is a collection of WSI tissues for the sentinel lymph node. It contains CAMELYON16 and CAMELYON17 challenges that include 399 WSI and 1000 WSI, respectively. The data are currently accessed via registration on the CAMELYON17 website [114]. The Breast Cancer Histopathological Image (BreakHis) contains 9109 macroscopic images for the tissue of the breast tumor obtained from 82 patients in various magnifying factors (40X, 100X, 200X). Up to now, it includes samples of 2480 benign and 5429 malignant WSIs [115].

Table 3. Some common downloadable WSI databases.

| Datasets | No Slides | Staining | Diseases |
|---|---|---|---|
| TCGA [33,116] | 18,462 | H&E | Cancer |
| GTE [112] | 25,380 | H&E | Normal |
| TMAD [113,117] | 3726 | H&E/IHC | various tissue |
| TUPAC16 [118] | 821 from TCGA | H&E | Breast cancer |
| Camelyon17 [114] | 1000 | H&E | Breast cancer (lymph node metastasis) |
| Köbel et al. [119,120] | 80 | H&E | Ovarian carcinoma |
| KIMIA Path24 [121] | 24 | H&E/IHC | various tissue |

Table 4. Some publicly available hand-annotated histopathological images.

| Datasets | No of Images | Staining | Organs | Potential Usage |
|---|---|---|---|---|
| KIMIA960 [122,123] | 960 | H&E/IHC | Different tissue | Classification |
| Bio-segmentation [124,125] | 58 | H&E | Breast | Classification |
| Bioimaging challenge 2015 [126] | 269 | H&E | Breast | Classification |
| GlaS [127] | 165 | H&E | Colorectal | Gland segmentation |
| BreakHis [115] | 7909 | H&E | Breast | Classification |
| Jakob Nikolas et al. [123,128] | 100 | IHC | Colorectal | Detection of blood vessel |
| MITOS-ATYPIA-14 [129] | 4240 | H&E | Breast | Detection of mitosis, classification |
| Kumar et al. [122,130] | 30 | H&E | Different cancer | Segmentation of Nuclear |
| MITOS [20] | 100 | H&E | Breast | Detection of mitosis classification |
| Janowczyk et al. [131,132] | 374 | H&E | Lymphoma | |
| Janowczyk et al. [131,132] | 85 | H&E | Colorectal | Segmentation of gland |

| Ma et al. [133] | 81 | IHC | Breast | TIL analysis |
| Linder et al. [134,135] | 1377 | IHC | Colorectal | Segmentation of epithelium and stroma |

## 6. Discussion and Histopathological Tasks

Since HI analysis is inherently a cross-disciplinary area, this review has stated that ongoing research is anticipated to have an obvious and tangible impact on automated HI analysis techniques. This paper reviews the recent state of the art CAD techniques for HI. This review also briefly describes the development of histopathology analysis and its problems. Recently, DL outperformed state-of-the-art techniques in various MLs for HI analysis tasks, such as recognition, classification, and segmentation. DL's merit, compared to other forms of learners, is their ability to acquire the performance as well or better than a human's performance. Currently, DL and WSI are revolutionizing the CAD of histopathology, and soon they could help reduce pathologists' workload in most simple tasks. This would allow pathologists to focus on challenging cases and lead to a deeper comprehension of pathologic procedures via ML techniques. More applications of HI analysis using ML techniques have been introduced in this review [111]. Most of the research developed in the field of HI analysis is addressed for some specific tasks.

*Tasks for Histopathology Image*

Open objective problems targeting issues in HI analysis were presented recently, such as in other medicinal imaging fields. A benefit of trying various methods on an unchanging dataset and in exact issues is the target comparison of advantages and constraints of literature methods. Especially in a sample of pathology, consuming time, and challenging issues of searching WSIs for appropriate tissue basics, such as nuclei and mitosis, could be enhanced. Choosing the most appropriate method to aid and advance the visible model of slides could help pathologists concentrate on significant issues when analyzing the mentioned studies. The difficulty of issues has improved recently. The objectives of problems could be gathered into three major issues, as shown in Figure 5.

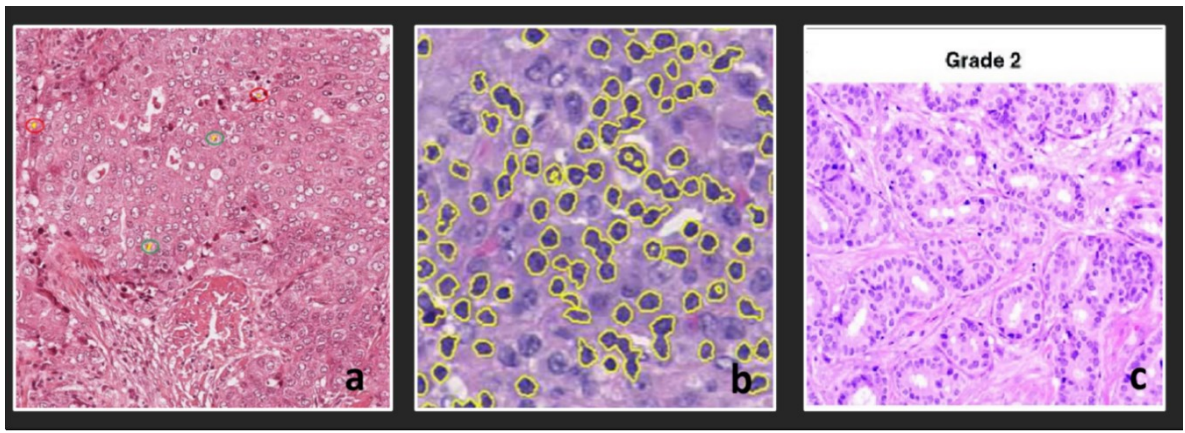

**Figure 5.** Challenges of HI analysis: (**a**) Mitosis detection, (**b**) Segmentation, and (**c**) Grade classification of a tumor.

- **Recognition of Mitosis**

Recognize mitosis contained in large power domains: There is a powerful relationship with the aggression of carcinoma and faster cell separation in extra mitosis. An essential part of HI tasks is the proper selection of evaluation metrics. A task of mitosis recognition utilizes the F1-score as the best metric to evaluate the participant techniques. The F1-score is calculated using Equation (10).

$$\text{F1-score} = \frac{2 \cdot precision \cdot recall}{precision + recall} \tag{10}$$

where Precision = TP/(TP + FP) and Recall = TP/(TP + FN). The threshold of maximum Euclidean distance from centroid for considering the mitotic event, as TP, was estimated to less than 7.5–8 μm.

- **Segmentation of structure**

The segmentation process localized and outlined the border of particular tissue architectures—for example, nuclei of cell or gland. Various kinds of tissue structures have been aimed to structure segmentation in HI. Automatic segmentation system output is typically evaluated by measuring some standard objective parameters, such as the mean boundary distance, the Dice coefficient, and Hausdorff Distance (HD). HD is one of the most insightful and useful metrics, since it measures the greatest segmentation error. For the two datasets, X and Y, the one-sided HD from X to Y is defined as:

$$\text{hd}(X;Y) = \max_{x \in X} \min_{y \in Y} \|x - y\|2 \tag{11}$$

and similarly, for hd (Y; X):

$$\text{hd}(Y;X) = \max_{y \in Y} \min_{x \in X} \|x - y\|2 \tag{12}$$

The bidirectional HD between these two sets is then:

$$\text{HD}(X;Y) = max(hd(X,Y), hd(Y,X)) \tag{13}$$

- **Classification of images**

First, one must find the characteristic set of features for a specific class of tissue, potentially taking into account the primitives of the underlying tissue. In histopathology, various evaluation techniques for ranking the classifications methods have been used. Various evaluation methods for rating the classification methods include

- For nuclear atypia rating, a points-based scheme is used.
- The region under the curve of the receiver operating characteristic (ROC) to classify slides of lymph node comprising metastasis or not.
- Approval with ground truth calculated with Spearman's correlation or quadratic weighted Cohen's kappa to grade cancer.

Equation (14) is used to calculate quadratic weighted Cohen's kappa.

$$K_w = 1 - \frac{\sum_{i,j} w_{i,j} \ p_{i,j}}{\sum_{i,j} w_{i,j} \ e_{i,j}} \tag{14}$$

$$e_{i,j} = \ p_{i,j} q_{i,j} \tag{15}$$

where $w_{i,j}$ represent weights, $p_{i,j}$ observed probabilities, and $q_{i,j}$ expected probabilities [136]. Though these issues were split into various sets due to their explanation, they could be mixed or regarded as a preliminary phase to another issue. For example, the WSI grading technique of carcinoma could begin by classifying the image as a tumor tissue. Next, the carcinoma was segmented. Finally, grade WSI dependent on the count of mitosis in the ROI of cancer.

**7. Limitations and Future Trends**

Digital HI recognition is an appropriate issue for ML because pictures themselves include data adequate for diagnosis. Issues in the analysis of digital HI applying ML is mentioned in this review. Because of reasonable efforts produced up to now, these issues being overcome, but there is space for enhancement. Many of these issues are probably resolved when a large amount of well annotated WSIs becomes obtainable. Collecting WSIs from different institutions to note them the exact conditions and creating this information public will be adequate to improve the growth of more advanced electronic HI analysis. Lastly, some possible future issues for the study are recommended, which have not been adequately researched.

- **Novel Objects Discovery**

For instance, unexpected items, irregular organization, uncommon tumor (not contained in the training stage), and aliens' bodies might exist in real diagnostic conditions. However, one can use a discrimination framework containing CNN classes, such as items among the predefined classes [137]. To solve the issue, the recognitions of outlier approaches were applied to HI. However, just a few studies have handled the issue up to now [138]. Recently, some DL-based techniques applied reconstruction error for recognition of outliers in other fields. However, they are not yet used in HI analysis.

- **Interpretable DL Model**

DL is usually disapproved of, since its decision-making process is not clear to individuals and thus frequently explained like a black box. People need to know the process of decision making or the basis of the decision. This might cause new findings in the domain of pathology. Even though this issue has not been fully resolved, some studies have tried to supply solutions, such as combined pathological pictures learning and diagnostic studies incorporated with interesting mechanisms [139]. In other fields, the basis of the decision might be ultimately displayed by visualizing the reaction of the deep network [140] or introducing a useful training picture applying impact functions [141].

- **Intraoperative Diagnosis**

Diagnosis by the pathologist during surgery impacts intraoperative decision making. Therefore, it might be another actual application in HI analysis. Because diagnosis time in an intraoperative examination is limited, a quick classifier while maintaining precision is significant. As a result of time limitation, the quick-freezing part is utilized rather than the formalin-fixed paraffin-embedded part that requires more time to get ready. Thus, for this reason, classification training must be executed, applying freezing part slides. Since the amount of appropriate WSI for analysis is not adequate, and function is more complicated than formalin-fixed paraffin-embedded slides, few studies have analyzed freezing parts [142].

- **Tumor-Infiltrating Immune Cell Analysis**

The microenvironment of carcinoma for immune cells has acquired significant interest recently. Thus, quantitative analysis for the carcinoma permeating of immune cells for slides applying ML methods is going to be among the emerging styles in HI analysis. Functions connected to analysis contain immune cell recognition in the H&E staining picture [143 ,144 ] and are recognized as a more specific form of immune cells applying immunohistochemistry [133]. Additionally, the structure of immune cell permeation and immune cell vicinity are supposedly linked to tumor treatment [145], spatial association analysis among immune cells and cells of cancer, and the association among this information and reaction to immunotherapy applying specific techniques, such as methods based on the graph [146], is likewise of good importance.

- **Challenges in HI analysis**

Typical DL architectures need their inputs in a particular structure with specific spatial dimensions. Moreover, these architectures are usually created for RGB pictures, while in digital HI, dealing with pictures in grayscale, HSV color might be desired for a particular system. Transforming pictures between color spaces, resizing pictures to suit GPU's storage, and determining the most effective resolution for applying at tilling are a few of the possible studies required, which will cause various levels of data loss. An acceptable information processing technique seeks to accomplish minimal data loss while using architectures for their maximal capacity. Input images are likely tiled or resized in most applications. It is also essential to balance the appropriate contexts and magnification with memory and computational constraints. Since CNN's can learn from smaller images more easily, images are not larger than the necessary context. A large amount of work was done to integrate low- and high-resolution inputs in different ways and issues to make better decisions [147].

- **Quality of training**

DL's accomplishment depends on the accessibility to high-quality training models to accomplish the required predictive efficiency [148,149]. Some efforts have been built to create extra annotated information by utilizing alternative methods, such as information augmentation [150], picture synthesis [107]. However, it is not even apparent that they are befitting from digital pathology.

- **Clinical translation**

There is a huge rapid development in AI research used in MI, and their possible effect has been shown by systems including the recognition of breast cancer metastasis [151], brain recognition [152], diagnosing diseases in retinal pictures [153], and so on. Regardless of this variety of systems, AI's actual and impactful implementation in medical practice will include many methods still to come.

- **Synthesis rather than marking**

An issue is that mapping of the label to the image domain is often unclear because the label mask can be mapped to many images. The training of the entire Generative Adversarial Network architecture can be difficult. The sizes of the regions-of-interest are given complexity here. Regions can display a diameter of up to several hundred pixels or thousands. This can be a big challenge, as the segmentation networks are implemented patch-wise.

- **Translation of morphology**

The optimal architecture for modified morphology settings does not show. Usually, unclear mappings can be particularly problematic in the event of morphological changes.

## 8. Conclusions

Different steps to analyze HIs are studied in this review for objective diagnosis automatically. In this survey, a comprehensive overview of different strategies in traditional and DL models has been presented. Different perspectives have tackled the analysis of HI for a wide variety of histology tasks (e.g., segmentation, tumor recognition, tissue classification). We have identified those that have been applied to various types of cancer (e.g., breast, kidney, colon, lung). For CAD in HI, there are primarily three phases: segmentation, feature extraction, and classification. The techniques developed for automatic analysis and evaluation of HIs help the pathologists in objective diagnosis for disease and decreased human error. A reference guide to recent literature methods for analyzing HIs manifests itself in the categorization techniques presented in this survey.

**Author Contributions:** Conceptualization, N.E. and M.E.; methodology, N.E., H.S., and M.E.; formal analysis, N.E. and M.E.; investigation, N.E., H.S., S.E.-S., S.M.R.I., and M.E.; writing—original draft preparation, N.E., H.S., and M.E.; writing—review and editing, N.E., H.S., S.E.-S., S.M.R.I., and M.E.; supervision, H.S. and M.E.; project administration, M.E.; All authors have read and agreed to the published version of the manuscript.

**Funding:** This research received no external funding.

**Conflicts of Interest:** The authors declare no conflict of interest.